Are Dark Energy and Dark Matter Different Aspects of the Same Physical Process?

Ruth E. Kastner[1] and Stuart Kauffman[2]

August 16, 2017

ABSTRACT. It is suggested that the apparently disparate cosmological phenomena attributed to so-called 'dark matter' and 'dark energy' arise from the same fundamental physical process: the emergence, from the quantum level, of spacetime itself. This creation of spacetime results in metric expansion around mass points in addition to the usual curvature due to stress-energy sources of the gravitational field. A recent modification of Einstein's theory of general relativity by Chadwick, Hodgkinson, and McDonald incorporating spacetime expansion around mass points, which accounts well for the observed galactic rotation curves, is adduced in support of the proposal. Recent observational evidence corroborates a prediction of the model that the apparent amount of 'dark matter' increases with the age of the universe. In addition, the proposal leads to the same result for the small but nonvanishing cosmological constant, related to 'dark energy,' as that of the causet model of Sorkin *et al*.

[1] University of Maryland, College Park; rkastner@umd.edu
[2] Institute for Systems Biology, Seattle and Professor Emeritus, Dept. of Biochemistry and Biophysics, University of Pennsylvania

1. Introduction.

Since the 1990s it has become clear that the universe is expanding at an accelerating rate, a phenomenon that was historically attributed to so-called "dark energy."[3] The hypothetical dark energy is invisible, and can be thought of as an intrinsic property of spacetime rather than usual matter (stress-energy) that is the source of spacetime curvature. The density of 'dark energy' is constant, also in contrast to ordinary matter/energy. A popular method of accounting for the dark energy phenomenon is by attributing it to Einstein's 'cosmological constant' $\Lambda$ (Einstein 1917).

An ostensibly separate phenomenon—the flattening of galactic rotation curves with radial distance--is also well known (e.g., Rubin et al 1980). This unexpectedly large value of rotational velocities for the outer observable matter in galaxies is an anomaly for standard Newtonian and Einsteinian gravitational theories, and in order to preserve them, it has been attributed to an invisible hypothetical form of matter dubbed 'dark matter.' However, rather than postulate 'dark matter,' some researchers have been exploring modifications of Newtonian gravitational theory. One such effort, "Modified Newtonian Dynamics" or MOND, was introduced by Milgrom (Milgrom, 1983). MOND has been successful in fitting the observed rotation curves, but it has the drawback of being an *ad hoc* alteration to the basic gravitational theory.

The situation has recently progressed significantly: Chadwick, Hodgkinson, and McDonald (2013) have proposed a modification of Einstein's general relativity based on the principle that (idealized) point masses give rise not only to the usual spacetime curvature, but also to spacetime expansion. For a particular value of the parameter governing how rapidly space expands, they find that their theory perfectly fits the galactic rotation data.

---

[3] E.g., Huterer, D.; Turner, M. (1999).

Currently, there is no known physical mechanism or process underlying the phenomena attributed to dark matter and dark energy (or the finite value of Λ if that is an accurate expression of the latter effect). This paper proposes such a physical process: a specific kind of spacetime emergence underlying a form of matter-based spacetime expansion that has not been previously taken into account. Thus, given the quantification of spacetime expansion by the CHM theory, we may be able to physically account for the "dark matter" phenomenon through a previously unsuspected expansion generated by ordinary matter. In addition, "dark energy" may be understood as an artifact of the same emergence process, arising from the discreteness of spacetime and its quantum origins.

In what follows, we first review the proposed model of spacetime emergence and then show that it naturally leads to the description provided by the CHM theory. Then we discuss another aspect of the emergence process that naturally leads to the nonvanishing, but very small, value of Λ that accounts for the 'dark energy' phenomenon.

2. Source of the spacetime expansion around mass points

The present authors have independently proposed that new elements of spacetime emerge from the quantum substratum through a real non-unitary process of measurement, in which quantum *potentiae* (a la Heisenberg, 1958) become actualized as new sets of structured spacetime events. One of us, REK, has proposed such a process of actualization and spacetime emergence as a key component of the relativistic extension of the Transactional Interpretation, now called the Relativistic Transactional Interpretation (RTI) (cf. Kastner 2012a, Chapter 8; Kastner 2012b).[4]

---

[4] An earlier, purely nonrelativistic version of TI originated by Cramer (1986) was subject to a challenge by Maudlin (2011, 184-5), but that has been completely nullified by the relativistic development resulting in RTI (Kastner 2017).

The other, SK, has independently been exploring the idea that measurement is a real physical process that converts quantum possibilities (understood as a new metaphysical category, *res potentia*) to spacetime actualities (identified as Descartes' *res extensa*) in the context of biophysics (Kauffman 2016, primarily Chapter 7). Both proposals, though having been arrived at and presented in different ways, lead to the same basic idea: spacetime expansion is always associated with 'measurement' at the quantum level, understood as a real (but inherently indeterministic) physical process.

In RTI, quantum objects, as described by quantum states, (solutions to the Schrodinger equation or, at the relativistic level, Fock States) are taken as elements of a quantum substratum that is a precursor to spacetime. That is, quantum objects are Heisenbergian *potentiae* (tokens of *res potentia* in Kauffman's terminology) that are not spacetime objects. They can be understood as necessary but not sufficient conditions for spacetime events. The transactional process (as detailed, for example, in Kastner 2012a, Chapter 3) is the sufficient condition that results in actualization of a spacetime interval $I$ as defined by an emission event $E$, an absorption event $A$, and the directed temporal and spatial connection between them, which is the transferred quantum (such as a photon). In this picture, energy and momentum are interpreted physically (not just mathematically) as the generators of temporal and spatial displacement, respectively.

Thus, a new spacetime interval $I(E,A)$ is physically generated as a result of a transaction: one which did not exist before. $I(E,A)$ is distinguishable in the sense that it has in-principle observable properties related to its identification with the process connecting $E$ and $A$ (e.g., energy and directional momentum transferred from $E$ to $A$.) An ongoing process of such transactional transfers from emitters and absorbers (i.e., atoms and molecules in the substratum, which can change roles from emitter to absorber and back again by repeatedly becoming excited and decaying) leads naturally to key aspects of the causal set ("causet") model of Sorkin et al (e.g., Rideout and Sorkin 2000 and references therein). However, in the RTI picture, each such spacetime event is contingent on the specific physical nature of the transaction that established it. This physically distinguishes and characterizes the spacetime events and their connections, so that they are not just generic 'atoms of spacetime' as in the causet model thus far.

More specifics regarding the process spacetime emergence in the RTI ontology is provided in Kastner (2016). It is shown in Kastner (2012a,b) that transactions (and thus new structured sets of spacetime events) occur with probabilities associated with decay rates, which are always Poissonian. Interestingly, Bombelli, Henson and Sorkin (2006) have independently found, with respect to the causet approach, that the growth of the causet in a Poissonian manner preserves Lorentz covariance.

The present proposal differs from that of Sorkin and his collaborators in that the spacetime substratum (i.e., the manifold that is the precursor to the spacetime causet) is comprised of specific quantum entities described by quantum states (i.e., field excitations that are created and destroyed). As noted above, these quantum entities stochastically give rise to new elements of the causet in a Poissonian process (Kastner 2016). In this picture, there are many possible (candidate) events for addition to the spacetime causet, but there is just one actual growing causet, and that is the emergent spacetime. The structure of that growing spacetime is contingent on the specific quantum entities (and their interactions) in the substratum; thus, it is those which will dictate the transition probabilities from a causet with N elements to a larger one with N+1 elements, rather than a transition probabilities applying to an arbitrary Markov process as in the classical sequential growth model (intended as a first step toward a quantum version of causet growth) studied in Rideout and Sorkin (2000).[5] Nevertheless, the fact that the uncertainty ∆N in the number of elements N is Poissonian leads to the same prediction for the cosmological constant as found by Sorkin et al, and therefore a physical basis for 'dark energy'; we turn to that in Section 3.

---

[5] Thus, in the RTI picture (as opposed to the approach of Sorkin et al), a theory of 'quantum gravity' consists of quantifying the correspondence between the elements of the quantum substratum and the emergent spacetime causet structure, the latter being the gravitational metric. A promising way forward in this regard is through the poset work of Knuth et al (e.g., Walsh and Knuth 2015).

How can we understand the new spacetime interval created in an actualized transaction as a form of spacetime expansion around a mass point, in order to find correspondence with the CHM theory accounting for "dark matter"? At the quantum level, a 'mass point' would be something like an isolated atom; say a hydrogen atom H. According to the current proposal, the atom is part of the quantum substratum—not a spacetime object—unless it is 'measured', i.e., engages in a transaction in terms of RTI. In order for H to count as a persistent mass point that could serve a source of stress-energy, it would have to be subject to ongoing measurement-- engaging in transactions that enable it to approximate a spacetime trajectory (see, e.g. Kastner 2012, Section 4.4).[6] These ongoing transactions (arising from other emitters and absorbers in the universe including Earth-based astronomical equipment) serve to repeatedly actualize H, and with every actualization, a new spacetime interval is created that did not exist before. This results in an observable expansion of the metric in the locus of H, in addition to any curvature already accounted for in standard general relativity.

According to this proposal, the process of spacetime expansion around matter should be an ongoing process, which leads to a specific prediction: the effect should increase monotonically with increasing proper time of the universe $\tau$. In fact, this effect has just recently been observed: very distant (i.e., large redshift, and therefore very young, recently born) galaxies have rotation curves much closer to the Newtonian gravitational prediction than do older galaxies (Genzel et al). (Of course, Genzel et al interpret the data based on the usual assumption that 'dark matter' really exists; they therefore tentatively conclude that the difference has to do with less 'dark matter' in the past in relation to the amount of normal baryonic matter.) We take this is a tentative corroboration of the model, but of course more observations are called for. In particular, it is now possible to study dark matter as a function of the age of a galaxy, and in addition, it may be possible to ascertain whether dark matter is spatially isotropic, or shows any variation with the density of observable matter.

---

[6] This process of a quantum system approximating a classical trajectory through measurement is well-known (not solely an aspect of RTI) and is related to the well-known 'inverse Zeno effect' (see, e.g., Panov 2001).

3. The cosmological constant and 'dark energy.'

We now return to the issue of 'dark energy'. As noted above, the result of the transactional spacetime emergence process is to yield a causal set of the sort contemplated by Sorkin et al, although the elements of the set have more structure in this picture; they are networked transactions $I(E_i,A_j)$ (where the indices are a shorthand representing birth order, chain membership, conserved physical quantities transferred, etc.[7]). In this regard, they more closely resemble the 'influence network' of Knuth et al (e.g., Knuth and Bahreyni 2014). Nevertheless, the fact that elements of causet are added in Poissonian fashion means that the current model yields the same nonvanishing, but very tiny, value for $\Lambda$.

Specifically, in natural units (h=G=1) $\Lambda$ has units of inverse length squared, and observations indicate that

$$\Lambda \lesssim 1/V^{1/2} \tag{1}$$

Based on empirical data, $\Lambda$ must be very close to zero; but to a first order approximation, one might find a very small but non-negligible value.[8] Sorkin (2007) provides such a first-order approximation, as follows. One notes (based on unimodular gravity[9]) that $\Lambda$ and V are essentially conjugate; i.e.,

$$\Delta\Lambda \, \Delta V \sim 1 \qquad (2)$$

---

[7] A 'chain' is a subset of a causet possessing a total order of its elements, providing a timelike relationship among them).
[8] For a discussion of the puzzle of small $\Lambda$, see Ng and van Dam (2001).

(in natural units), analogously to the quantum mechanical uncertainty relations. Sorkin notes that this conjugate relationship between $\Lambda$ and V is evident from the action integral,

$$S = -\Lambda \int (-g)^{1/2} d^4x = -\Lambda V \qquad (3)$$

Thus, if $\Lambda$ is to have any non-vanishing value, it must be due to its uncertainty

$$\Delta\Lambda \sim 1/\Delta V \qquad (4)$$

based on any uncertainty in V. In the causet model, V is proportional to N, since the latter specifies how many 'atoms of spacetime' exist; or, in the RTI picture, how many $I(E_i, A_j)$ have been actualized. Now, given that elements are added to the (discrete) spacetime manifold in a Poissonian process, the number N of elements has an intrinsic uncertainty of $N^{1/2}$ for any given value of the proper time $\tau$. Since V is a function of $\tau$, V inherits this uncertainty: $\Delta V \sim V^{1/2}$. If the uncertainty is the only (significant) contribution to the value of $\Lambda$, then we get precisely (1).

A more direct way to get the result (1) is through (3), which shows that the action $S = \Lambda V$. Sorkin observes that $\Lambda = S/V \approx S/N$ (modulo units based on a fundamental spacetime 'length,' $l = (\hbar \, 8\pi G/c^3)^{1/2}$), saying that $\Lambda$ "can be interpreted as the action per causet element that is present even when the spacetime curvature vanishes. As one might say, *it is the action that an element contributes just by virtue of its existence.*" (Sorkin 2007, p. 8; emphasis added.) He goes on to observe that in a random process (associated with error √N), wherein each element comprising the causet contributes $\pm\hbar$ to S, one obtains

$$S \sim \pm\hbar\sqrt{N} = \pm\hbar\sqrt{\frac{V}{l^4}} \qquad (5)$$

and therefore, from (3):

$$\Lambda = S/V = \pm\frac{\hbar}{l^2\sqrt{V}} \qquad (6)$$

which gives us a more precise form of (1).

---

[9] I.e., the condition that the metric tensor *g* has unit determinant.

This line of reasoning is presented as an 'ansatz' in Sorkin (2007), but it has a more direct physical grounding in RTI, in which spatiotemporal displacements are generated through transfers of 4-momentum. In this picture, the uncertainty relation describes the fact that it takes a finite time Δt to transfer (actualize) a well-defined quantity of energy ΔE, and a finite spatial distance Δx to transfer (actualize) a well-defined quantity of momentum Δp. Thus, each spacetime interval actualized in a transaction that transfers a quantity of 4-momentum from emission event *E* to absorbing event *A* is indeed physically characterized by a quantum of action of magnitude $\hbar$. Thus, we gain a physical basis for Sorkin's insightful ansatz above. Sorkin adds a caveat in a footnote; but we believe he was right in the first place, and that allowing for the emergence of complete spacetime intervals characterized by the unit of action as the basic, indivisible 'atoms of spacetime' places this insight on sound physical footing.

4. Conclusion

We have shown that a specific mechanism of spacetime emergence from the quantum level leads to the spacetime expansion quantitatively described in the theory of Chadwick, Hodgkinson, and McDonald (2013), which correctly predicts observed galaxy rotation data attributed to 'dark matter.' In addition, we have shown that the same mechanism yields a discrete spacetime characterized by Poissonian uncertainties, similar to that proposed by Sorkin et al, which results in the necessary value of Λ to account for the 'dark energy' phenomenon, according to current observational data. In this model, we may understand 'dark energy' as a property arising from each element of spacetime, as Sorkin says, *"just by virtue of its existence"*—since a finite quantity of action is required in order for it to exist.

This possible relation of dark energy and matter is intriguing, as it would unify apparently disparate and yet equally unexpected cosmological phenomena. If an expansion of spacetime around mass points can account for the excess rotation of the outskirts of galaxies (i.e., "dark matter"), and if this expansion is related to dark energy as outlined herein, we gain explanatory parsimony as well as evidence for a fascinating connection of spacetime with the quantum level. The latter could aid efforts to find a theory of quantum gravity.

Acknowledgments: The authors are grateful to Jim Bogan for a critical reading of the manuscript and for catching a typo.